\documentclass[useAMS,twocolumn,usenatbib]{mn2e}

\usepackage{graphicx,enumitem}

\usepackage{amssymb,amsmath}

\newcommand{\myvec}[1]{\boldsymbol{#1}}

\begin{document}

\title[MG-GADGET: A new code for modified gravity simulations]
{Modified Gravity-GADGET: A new code for cosmological hydrodynamical simulations of modified gravity models}
\author[Ewald Puchwein, Marco Baldi, Volker Springel]
{Ewald Puchwein$^{1}$, 
Marco Baldi$^{2,3}$  and
Volker Springel$^{1,4}$
\\$^1$Heidelberger Institut f{\"u}r Theoretische Studien, Schloss-Wolfsbrunnenweg 35, 69118 Heidelberg, Germany
\\$^2$Dipartimento di Fisica e Astronomia, Universit\`a di Bologna, Viale B.~Pichat 6/2, 40127 Bologna, Italy
\\$^3$INAF, Osservatorio Astronomico di Bologna, Via Ranzani 1, 40127 Bologna, Italy
\\$^4$Zentrum f\"ur Astronomie der Universit\"at Heidelberg, Astronomisches Recheninstitut, M\"{o}nchhofstr. 12-14, 69120 Heidelberg, Germany}
\date{\today}
\maketitle

\begin{abstract} 
  We present a new massively parallel code for N-body and cosmological
  hydrodynamical simulations of modified gravity models. The code
  employs a multigrid-accelerated Newton-Gauss-Seidel relaxation
  solver on an adaptive mesh to efficiently solve for perturbations in
  the scalar degree of freedom of the modified gravity model. As this
  new algorithm is implemented as a module for the {\sc p-gadget3} code, it
  can at the same time follow the baryonic physics included in {\sc
    p-gadget3}, such as hydrodynamics, radiative cooling and star
  formation. We demonstrate that the code works reliably by applying
  it to simple test problems that can be solved analytically, as well
  as by comparing cosmological simulations to results from the
  literature. Using the new code, we perform the first non-radiative
  and radiative cosmological hydrodynamical simulations of an
  $f(R)$-gravity model. We also discuss the impact of AGN feedback on
  the matter power spectrum, as well as degeneracies between the
  influence of baryonic processes and modifications of gravity.
\end{abstract}

\begin{keywords}
cosmology: theory -- methods: numerical
\end{keywords}

\section{Introduction}
\label{sec:introduction}

One of the most puzzling open issues in modern physics concerns the
origin of the observed late-time accelerated expansion of the Universe
\citep[][]{Riess_etal_1998,Perlmutter_etal_1999,Schmidt_etal_1998}.
This acceleration has no natural explanation in the framework of a
cosmological model based on Einstein's General Relativity and standard
particle physics. Therefore, either new fields with exotic global
properties (as, e.g., a negative effective pressure) or new physics in
the gravity sector have to be invoked in order provide a consistent
description of the cosmic evolution.  The former approach is generally
referred to as ``Dark Energy" (DE, hereafter), while the latter goes
under the name of ``Modified Gravity" (MG).

More conservative explanations for the observed cosmic acceleration
have also been attempted, for example as a backreaction effect of the
formation of nonlinear cosmic structures on the background expansion
of the universe \citep[see
  e.g.][]{Kolb_Matarrese_Riotto_2006,Rasanen_2011,Clarkson_etal_2011}
or through the presence of a very large void surrounding the Milky
Way's location and mimicking a late-time accelerated expansion
\citep[see
  e.g.][]{Mustapha_Hellaby_Ellis_1997,Tomita_2001,Wiltshire_2007}.
However, such possibilities have been claimed either to produce an
insufficient effect on the cosmic evolution to account for the
observed acceleration \citep[see e.g.][]{Behrend_Brown_Robbers_2008,Green_Wald_2011,Baumann_etal_2012}, or to be at odds with
independent observational data \citep[as shown
  e.g. by][]{Zumalacarregui_Garcia-Bellido_Ruiz-Lapuente_2012}.

The presently accepted cosmological model \citep[e.g.][]{Planck_016} -- where the accelerated
expansion is driven by a cosmological constant $\Lambda $ --
represents the boundary between standard physics and the wide range of
DE and MG extended scenarios. Any extension of particle physics or of General Relativity besides a
cosmological constant necessarily introduces additional degrees of
freedom that would indicate the incompleteness of one or both of these
two pillars of modern physics. It is therefore not surprising that the
detection of possible deviations from the expected behaviour of a
cosmological constant has become a primary target for a number of
large and challenging observational campaigns, such as e.g. PanStarrs
\citep[][]{PanStarrs}, HETDEX \citep[][]{HETDEX}, DES \citep[][]{DES},
LSST \citep[][]{LSST} and Euclid\footnote{www.euclid-ec.org}
\citep[][]{EUCLID-r}.

Moving beyond the boundary between standard and non-standard physics
represented by the cosmological constant corresponds to opening a
Pandora's box in terms of possible alternative and competing theories
\citep[see e.g. ][for a recent comprehensive review]{Euclid_TWG}.
These include a large number of different realisations of dynamical DE
models, generally characterised by the cosmic evolution of a classical
scalar field, as (just to cite the most popular) {\em Quintessence}
\citep[][]{Wetterich_1988,Ratra_Peebles_1988}, {\em k-essence}
\citep[][]{kessence}, {\em Phantom} \citep[][]{Caldwell_2002} and {\em
  Quintom} \citep{Feng_Wang_Zhang_2005} DE scenarios, generally
assuming negligible spatial perturbations and interactions of the DE
scalar field. More sophisticated models can be obtained by dropping
such assumptions and allowing the DE scalar field to cluster at
sub-horizon scales and to have direct interactions with standard
particles as for the case of {\em Clustering Dark Energy}
\citep[][]{Creminelli_etal_2009,Sefusatti_Vernizzi_2011,Batista_Pace_2013}
and {\em Coupled Quintessence}
\citep[][]{Wetterich_1995,Amendola_2000,Farrar2004,Amendola_Baldi_Wetterich_2008,Baldi_2011a}.
On the other hand, several possible modifications of General
Relativity have also been proposed, including {\em Scalar-Tensor
  theories} of gravity \citep[as e.g. $f(R)$ models,
][]{Buchdahl_1970,Starobinsky_1980,Hu2007,Sotiriou_Faraoni_2010}, the
{\em DGP} scenario \citep[][]{Dvali_Gabadadze_Porrati_2000}, or the
{\em Galileon} model \citep[][]{Nicolis_Rattazzi_Trincherini_2009}.

All these MG theories, however, need to rely on specific screening
mechanisms capable of suppressing the deviations from the behaviour of
standard gravity in the local environment of our Galaxy in order to
comply with the very stringent observational tests of general
relativity that have been performed on Earth and within the solar
system \citep[see
  e.g.][]{Bertotti_Iess_Tortora_2003,Will_2005}. Several possible ways
to obtain an environment-dependent screening of the effects of a
modified theory of gravity have been proposed in recent years,
including the {\em Chameleon} \citep[][]{Khoury_Weltman_2004}, the
{\em Dilaton} \citep[][]{Gasperini_Piazza_Veneziano_2002}, the {\em
  Symmetron} \citep[][]{Hinterbichler_Khoury_2010} or the {\em
  Vainshtein} \citep[][]{Vainshtein_1972,Deffayet_etal_2002}
mechanisms. All such mechanisms are based on the development of large
(i.e. possibly nonlinear) spatial perturbations of the new degrees of
freedom associated to local matter overdensities, thereby recovering
standard gravity in high-density environments such as in our Galaxy.

Due to such nonlinearities, a fully consistent description of these
screening mechanisms and of their effects on structure formation needs
to rely on numerical techniques, and in particular on suitable
modifications of N-body algorithms capable of solving for the spatial
modulation of the deviation from standard gravity along with the usual
evolution of matter density perturbations \citep[see][for a recent
  review on the implementation of DE and MG models in N-body
  simulations]{Baldi_2012b}. These requirements have led in the last
years to the development of a few specifically-designed N-body codes
for MG cosmologies, starting from the first $f(R)$ algorithm realised
by \citet{Oyaizu2008} using a fixed-resolution particle-mesh approach,
to more recent implementations exploiting Adaptive Mesh Refinement
(AMR) techniques that were also extended to a wider variety of MG
models with different types of screening mechanisms \citep[see
  e.g][]{Khoury_Wyman_2009,Li2012,Brax_etal_2011,Davis_etal_2012,Llinares_Mota_2013}.
These various codes have been successfully employed to investigate
different possible observational features of MG theories in the
statistical and structural properties of collapsed CDM halos, as their
impact on e.g. the matter power spectrum \citep[see
  e.g.][]{Oyaizu2008,Schmidt_etal_2009,Li2012}, redshift-space
distortions \citep[][]{Jennings_etal_2012}, geometrical and dynamical
properties of virialized halos
\citep[][]{Lombriser_etal_2012,Lee_etal_2012,Lam_etal_2012,Llinares_Mota_2012},
or the statistics of CDM halos and voids \citep[as
  in][]{Zhao_etal_2010,Zhao_Li_Koyama_2011a,Winther_Mota_Li_2011,Li_Zhao_Koyama_2011}. These
numerical investigations have shown for a number of realisations of MG
cosmologies that they can result in sizeable effects on, e.g., the
matter power spectrum, the abundance of massive halos, or the
large-scale velocity field, at length and mass scales that will become
directly testable with the next generation of wide-field galaxy
surveys, with a particular relevance for Euclid and LSST.

The same scales and mass ranges, however, appear to be heavily
affected by a number of baryonic astrophysical processes \citep[see
  e.g.][]{Puchwein_etal_2005,Puchwein2008,Stanek_Rudd_Evrard_2009,VanDaalen2011,Semboloni2011,Casarini_etal_2011}
that have not been fully understood yet and whose implementation in
cosmological hydrodynamical simulations requires the development of
effective recipes to reproduce the integrated effects of physical
processes below the resolution limit. The implied degeneracy between
the observational features of cosmological models for the accelerated
expansion of the universe and astrophysical processes at small scales
represents one of the most problematic aspects for a full exploitation
of the wealth of high-quality data that will become available over the
next decade.  It is therefore particularly important to be able to
quantify the level of this degeneracy, or in other words to be able to
predict to which extent our assumptions on astrophysical processes at
small scales can bias our conclusions on cosmological model selection,
and vice versa.  To this end, the possibility to combine, in a single
cosmological simulation code, both the implementation of specific MG scenarios and
state-of-the-art prescriptions for small-scale baryonic processes,
like feedback from AGN, represents a crucial step in the development
of a robust interpretative framework within which to process and
analyse present and future data.

In this work, we present a new tool for cosmological N-body and
hydrodynamical simulations, called {\sc mg-gadget} that was developed
with the main goal of allowing such combined investigation of MG
theories and baryonic effects. It is implemented as a module for the
widely used Tree-PM SPH-code {\sc p-gadget3} and therefore inherits
its up to date implementations of gas cooling, star formation,
supernova and AGN feedback mechanisms, that have been developed in the
last years for this code. The MG solver is based on a
Newton-Gauss-Seidel relaxation scheme with adaptive resolution and
provides an efficient computation of the MG extra degrees of
freedom. In the remainder of the paper, we will describe the specific
MG scenarios that can be presently treated with our code
(Sec.~\ref{sec:methods}), we will discuss in detail the algorithm
implemented in {\sc mg-gadget} for the MG solver
(Sec.~\ref{sec:simulation_code}), as well as results of simple test
problems to assess the accuracy of the code
(Sec.~\ref{sec:tests}). Then we will present the first cosmological
simulations performed with {\sc mg-gadget} with a focus on the impact
of MG on the matter power spectrum (Sec.~\ref{sec:cosmo_runs}) and the
outcomes of some radiative and non-radiative hydrodynamical
simulations of MG models (Sec.~\ref{sec:cosmo_hydro_runs}). Finally,
in Sec.~\ref{sec:conclusions} we will summarize our results and draw
our conclusions.

\section{f(R)-gravity}
\label{sec:methods}

We will use $f(R)$-gravity models as a testbed and first application
of our MG code. Such models are generalizations of Einstein's general
relativity, i.e. the Ricci scalar $R$ in the Einstein-Hilbert action
is replaced by a function of $R$, or more precisely by $R+f(R)$ in our
notation. The action is then given by
\begin{equation}
  S = \int d^4x \, \sqrt{-g} \left( \frac{R+f(R)}{16 \pi G} + {\cal L}_m \right),
\end{equation}
where $G$ is the gravitational constant, ${\cal L}_m$ is the
Lagrangian density of matter and $g$ is the determinant of the metric
tensor. The field $f_R \equiv \frac{df(R)}{dR}$ can be interpreted as
a scalar degree of freedom of such models. For viable models with $f_R
\ll 1$ and in the quasistatic approximation, i.e. assuming
$|\myvec{\nabla} f_R| \gg \frac{\partial f_R}{\partial t}$, the field
equation for $f_R$ is given by \citep[see e.g.][]{Oyaizu2008}
\begin{equation}
  \nabla^2 f_R = \frac{1}{3c^2}\left(\delta R - 8 \pi G \delta \rho \right),
\label{eq:fR_field_eq}
\end{equation}
where we have here restored factors of the speed of light $c$, and
$\delta \rho$ and $\delta R$ are the perturbations in the density and
scalar curvature respectively. The gravitational potential $\phi$
satisfies \citep{Hu2007}
\begin{equation}
  \nabla^2 \phi = \frac{16 \pi G}{3} \delta \rho - \frac{1}{6}\delta R.
\label{eq:phi_poisson_eq}
\end{equation}

In the model described by \citet{Hu2007}, $f(R)$ is given by
\begin{equation}
  f(R) = -m^2 \frac{c_1 \left(\frac{R}{m^2}\right)^n}{c_2 \left(\frac{R}{m^2}\right)^n + 1},
\end{equation}
where $ m^2 \equiv H_0^2 \Omega_m $. Here, $H_0$ is the present day
Hubble parameter and $\Omega_m$ is the mean matter density in units of
the critical density of the Universe. $n$, $c_1$ and $c_2$ are free
parameters. When suitably chosen, this model exhibits the chameleon
mechanism, so that standard general relativity theory is restored in
massive objects, or more precisely within deep potential wells. A cosmic
expansion history that is close to $\Lambda$CDM can be obtained if we
require $c_2 (R/m^2)^n \gg 1$. $f(R)$ is then given by $f(R) = -m^2
c_1 / c_2 + O((m^2/R)^n)$. Setting the zeroth order term $-m^2 c_1 /
c_2$ equal to $-2\Lambda$, where $\Lambda$ is the desired cosmological
constant, then the model closely recovers the expansion history of
$\Lambda$CDM. This can be rephrased in a relation between the
parameters $c_1$ and $c_2$, namely
\begin{equation}
  \frac{c_1}{c_2} = 6 \frac{\Omega_\Lambda}{\Omega_m},
\label{eq:c1c2_lambda}
\end{equation} 
where $\Omega_\Lambda$ is the vacuum energy density in units of the
critical energy density of the Universe. The field
$f_R=\frac{df(R)}{dR}$ takes the form
\begin{equation}
  f_R = -n \frac{c_1 \left(\frac{R}{m^2}\right)^{n-1}}{\left[c_2
      \left(\frac{R}{m^2}\right)^n + 1\right]^2},
\end{equation}
in this model. Again assuming $c_2 (R/m^2)^n \gg 1$, this reduces to
\begin{equation}
  f_R \approx -n \frac{c_1}{c_2^2}\left(\frac{m^2}{R}\right)^{n+1}.
\label{eq:fR-R,n_relation}
\end{equation}

In a Friedmann-Robertson-Walker universe the scalar curvature is given by
\begin{equation}
  \bar{R} = 12 H^2 + 6 \frac{dH}{d \ln a} H,
\end{equation}
where $a$ is the scale factor and $H \equiv \dot{a}/a$ is the Hubble
function. For a flat $\Lambda$CDM expansion history this can be
rewritten as
\begin{equation}
  \bar{R}(a) = 3 \, m^2 \left( a^{-3} + 4 \frac{\Omega_\Lambda}{\Omega_m} \right).
\end{equation}
Evaluating this at $a=1$ and plugging the result into
Eq.~(\ref{eq:fR-R,n_relation}) yields an equation for $\bar{f}_{R0}$,
the background value of $f_R$ at $z=0$. Fixing the value
$\bar{f}_{R0}$, hence, results in a relation between $c_1$ and
$c_2$. Together with Eq.~(\ref{eq:c1c2_lambda}), this completely
determines both $c_1$ and $c_2$. It is, therefore, possible to
parametrise $f(R)$-models of this type by $n$ and $\bar{f}_{R0}$
rather than by $n$, $c_1$ and $c_2$. We adopt this convention
throughout the remainder of this paper. Furthermore, all the
simulations we present here assume $n=1$. In this case,
$\bar{f}_{R}(z)$ and $\delta R$ are given by
\begin{align}
  \bar{f}_{R}(a) &= \bar{f}_{R0}
  \left(\frac{\bar{R}_0}{\bar{R}(a)}\right)^2 = \bar{f}_{R0}
  \left(\frac{1 + 4 \frac{\Omega_\Lambda}{\Omega_m}}{ a^{-3} + 4
    \frac{\Omega_\Lambda}{\Omega_m}}\right)^2, \label{eq:fR_mean}
  \\ \delta R &= \bar{R}(a)\left(\sqrt{\frac{\bar{f}_{R}(a)}{f_{R}}} -
  1\right). \label{eq:dR_fR}
\end{align}
Together with Eqs.~(\ref{eq:fR_field_eq}) and
(\ref{eq:phi_poisson_eq}), this defines the set of equations that we
need to solve numerically for the force computation in our
cosmological simulations. Once these equations are solved, the
gravitational acceleration can be computed in the standard way from
the gradient of the gravitational potential, i.e. $\myvec{a} = -
\myvec{\nabla}\phi$.

Alternatively, combining Eqs.~(\ref{eq:fR_field_eq}) and (\ref{eq:phi_poisson_eq}) yields
\begin{equation}
 \nabla^2 \phi = 4 \pi G \delta \rho - \frac{c^2}{2} \nabla^2 f_R.
\end{equation}
From this, it is straightforward to identify the modified gravity
contribution to the gravitaional acceleration. It is given by
\begin{equation}
 \myvec{a}_{MG} = \frac{c^2}{2} \myvec{\nabla} f_R.
\label{eq:modgrav_accel}
\end{equation}
and can, hence, be directly computed from $f_R$.

\section{The simulation code}
\label{sec:simulation_code}

Our new fully parallel cosmological simulation code for modified
gravity models {\sc mg-gadget} (Modified-Gravity-{\sc gadget}) is
based on the TreePM-SPH simulation code {\sc p-gadget3}, an updated
and significantly extended version of {\sc gadget-2}
\citep{Springel2005c}. In order to use the code to follow structure
formation in modified gravity models, it was, obviously, necessary to
significantly change {\sc p-gadget3}'s gravity solver. All the changes
we implemented are described in detail below.

\subsection{Using the gravitational tree as an adaptively refined mesh}
\label{sec:mesh}

{\sc p-gadget3} uses a long-range/short-range force splitting to
calculate Newtonian gravitational forces (see \citet{Springel2005c}
for full details). Long-range forces are computed with Fast Fourier
transform methods, while short-range forces are evaluated using a
hierarchical octree \citep[see also][]{Barnes1986}. Such a scheme is
called a tree particle-mesh (TreePM) gravity algorithm. On coarse tree
levels, the nodes of this gravitational octree exactly tessellate the
simulation volume. On fine tree levels, instead, the nodes are no
longer space-filling but cover only high density regions. Depending on
how exactly the tree is built, the leaf nodes may not even be space
filling in these regions. However, by selecting nodes form different levels it
is possible to obtain a mesh that exactly tessellates the
whole simulation volume and has higher resolution in high-density
regions. In other words, nodes can be chosen that define an adaptive
mesh that refines on high-density regions. This adaptive mesh can then
be used to compute and store any scalar degree of freedom that the
modified gravity model under consideration has. In particular, we will
use it in the following to solve Eq.~(\ref{eq:fR_field_eq}) for $f_R$,
as well as for storing the corresponding effective mass density
(defined below in Eq.~(\ref{eq:effective_density})). The methods used
for solving for $f_R$ on this adaptive mesh are conceptually similar
to those employed by \citet{Li2012}. Full details will be given in
Secs.~\ref{sec:discrete_eqns} and \ref{sec:multigrid}.

To obtain the adaptive mesh, we start with the root node and
recursively check all tree nodes whether they should be selected as
AMR cells. The exact criteria for selecting a node are as follows:
\begin{enumerate}[itemindent=7mm]
  \item the node is not part of a coarser level node that has already
    been chosen as an AMR cell,
  \item the node has less than 8 daughter nodes, i.e. the daughter
    nodes (if they exist at all) do not cover the volume of the node
    completely and therefore would not qualify as AMR cells themselves.
\end{enumerate}
Using these criteria for a given tree yields the highest resolution
adaptive mesh that is space-filling. This is, however, not necessarily
the best choice as using cells that are much smaller than the
gravitational softening may not make sense. We, therefore, allow to
set a maximum tree level on which AMR cells may reside. Then, nodes
that are on this level and satisfy the first of the above criteria are
selected even if they have 8 daughter nodes. This maximum AMR level is
typically chosen such that the side length of nodes on this level is
somewhat smaller then the gravitational softening.
 
Above we have described how we select tree nodes as AMR cells, we
have, however, not yet specified how the tree is built in the first
place. {\sc p-gadget3} constructs the gravitational tree by
sequentially inserting particles into the tree. Whenever a leaf node
has already another particle attached in the octant in which the new
particle should be added a subnode is created in that octant. Both
particles are then attached to the subnode (unless both particles also
fall in the same octant of the subnode, in that case a sub-subnode
needs to be created). In this way, a sparse memory-efficient octree is
built. Nodes in this tree may have any number of subnodes between 0
and 8. The criteria specified above allow selecting nodes that define
a space-filling adaptive mesh even from such a sparse tree. However,
for unfavourable particle configurations the number of particles per
AMR cell can become quite large (up to several hundred particles). In
such cases, the code would then achieve a significantly worse
resolution on the AMR grid compared to the particle distribution. To
avoid this, one can add missing subnodes to AMR nodes that contain
many particles so that its daughter nodes qualify as AMR nodes
themselves. While one could think of different criteria for adding
subnodes, we used the most rigorous approach throughout the remainder
of this paper, i.e. whenever a new subnode of some node is created, we
also create the 7 other subnodes. Using this approach each AMR cell
contains at most 8 particles and typically $\lesssim 1$.

Before we can start solving for any scalar degree of freedom, we have
to assign masses to the AMR cells. We do this with the cloud-in-cell
(CIC) scheme, i.e. we represent each simulation particle by a constant
density cube and assign to each AMR cells that overlaps with the
cube the fraction of the particle mass that corresponds to the
overlapping fraction of the volume. To this end, the side length of
the AMR node that the particle resides in is adopted as the side
length of the cube. When applied in this way the CIC scheme can also
be used in regions in which the AMR grid changes resolution. Once a
CIC mass has been assigned to all AMR cells we also compute masses for
all coarser cells by simply summing up the masses of their daughter
cells.

\subsection{Discretisation of the f(R)-equations}
\label{sec:discrete_eqns}

Models that produce cosmic structures that are compatible with
observations correspond to small negative values of $\bar{f}_{R0}$. It
then follows by Eq.~(\ref{eq:fR_mean}) that $\bar{f}_R(a)$ is also
negative for all $a$. Eq.~(\ref{eq:dR_fR}) would then yield an
unphysical imaginary $\delta R$ if $f_R$ would attain a positive value
at some position. A physical solution for $f_R$ should, instead, be
negative everywhere and at all times. $f_R$ may, however, reach values
very close to zero. This is problematic when using an iterative
numerical solver to find $f_R$ as finite step sizes may at some point
yield a positive $f_R$ value. The resulting imaginary $\delta R$ would
then prevent us form continuing the iteration and, hence, make the
code unstable.

Following \citet{Oyaizu2008}, we do not solve directly for $f_R$ but
introduce a new field $u$ which we define by
\begin{equation}
  u \equiv \ln \left(\frac{f_R}{\bar{f}_R(a)}\right),
\end{equation}
so that
\begin{equation}
  f_R = \bar{f}_R(a) \times e^u.
\label{eq:fR_u}
\end{equation}
Using the field $u$ does by construction not permit $f_R$ to change
its sign and therefore avoids the above mentioned stability
issue. Using Eq.~(\ref{eq:dR_fR}), Eq.~(\ref{eq:fR_field_eq}) can then
be rewritten in terms of $u$ as
\begin{equation}
  \nabla^2 e^u + \frac{1}{3 c^2 \bar{f}_R(a)} \left[ \bar{R}(a) \left(
    1 - e^{-\frac{u}{2}} \right) + 8 \pi G \delta \rho \right] = 0.
\label{eq:u_field_eq}
\end{equation}
In order to solve this equation for $u$ on a grid, we need to
discretise $\nabla^2 e^u$. We do this in the same way as in
\citet{Oyaizu2008}, i.e.
\begin{align}
(\nabla^2 e^u)_{i,j,k} = 
  \frac{1}{h^2} \big[& b_{i-\frac12,j,k} u_{i-1,j,k} + b_{i+\frac12,j,k} u_{i+1,j,k} \nonumber \\ 
                     &- (b_{i-\frac12,j,k} + b_{i+\frac12,j,k}) u_{i,j,k}\big] \nonumber \\
+ \frac{1}{h^2} \big[& b_{i,j-\frac12,k} u_{i,j-1,k} + b_{i,j+\frac12,k} u_{i,j+1,k} \nonumber \\
                     &- (b_{i,j-\frac12,k} + b_{i,j+\frac12,k}) u_{i,j,k}\big] \nonumber \\
+ \frac{1}{h^2} \big[& b_{i,j,k-\frac12} u_{i,j,k-1} + b_{i,j,k+\frac12} u_{i,j,k+1} \nonumber \\
                     &- (b_{i,j,k-\frac12} + b_{i,j,k+\frac12}) u_{i,j,k} \big],
\end{align}
where $i,j,k$ are the cell indices in the $x$, $y$, and $z$
directions, respectively. $h$ is the physical side length of a cell
and
\begin{align}
  b_{i-\frac12,j,k} \equiv \frac 12 \left( e^{u_{i-1,j,k}} + e^{u_{i,j,k}} \right), \\
  b_{i+\frac12,j,k} \equiv \frac 12 \left( e^{u_{i,j,k}} + e^{u_{i+1,j,k}} \right).
\end{align}
The coefficients for the $y$ and $z$ directions,
i.e. $b_{i,j-\frac12,k}$, $b_{i,j+\frac12,k}$, $b_{i,j,k-\frac12}$,
and $b_{i,j,k+\frac12}$, are defined in an analogous way. We
additionally define
\begin{align}
  {\cal L}_{i,j,k} \equiv (\nabla^2 e^u)_{i,j,k} 
  + \frac{1}{3 c^2 \bar{f}_R(a)} \bar{R}(a) \left( 1 - e^{-\frac{u_{i,j,k}}{2}} \right)
\end{align}
and 
\begin{align}
  f_{i,j,k} \equiv \frac{1}{3 c^2 \bar{f}_R(a)} 8 \pi G \left(\bar{\rho}(a)-\frac{m_{i,j,k}}{h^3}\right),
\end{align}
where $m_{i,j,k}$ is the mass assigned to cell $(i,j,k)$ and
$\bar{\rho}(a) = \bar{\rho}_0 / a^3$ is the mean physical matter
density as a function of the scale factor. This allows us to
discretise Eq.~(\ref{eq:u_field_eq}) in the following form
\begin{equation}
  {\cal L}_{i,j,k} = f_{i,j,k}.
\label{eq:u_discrete_field_eq}  
\end{equation}
In the next section we will describe how to iteratively solve this equation.

\subsection{Iterative solver with multigrid acceleration}
\label{sec:multigrid}

Because of its nonlinearity, standard FFT-based methods are not
suitable for solving the field equation for $u$. Therefore, most
previous works \citep[e.g.][]{Oyaizu2008,Li2012} resorted to iterative
Newton-Gauss-Seidel relaxation schemes. We also follow this approach
here. Starting with an initial guess $u_{i,j,k,}^0$, we obtain a new,
typically more accurate approximation of $u$ by
\begin{equation}
  u_{i,j,k}^{n+1} = u_{i,j,k}^{n} - \frac{{\cal L}_{i,j,k}^n - f_{i,j,k}}{\frac{d {\cal L}_{i,j,k}^n}{d  u_{i,j,k}^{n}}}.
\label{eq:iterative_correction}
\end{equation}

To compute ${\cal L}_{i,j,k}^n$ and its derivative we need the value
of $u$ in cell $(i,j,k)$, as well as in its six direct neighbours. In
each iteration, we update the cells using a {\it red-black sweep},
i.e. using the three-dimensional equivalent of a chessboard
pattern. In the first half-sweep all {\it red} cells and in the second
all {\it black} cells are updated. In each half-sweep, a cell is,
thus, only affected by its six direct neighbours. The effect of some
source term, e.g. a point mass, on the approximate solution,
therefore, only propagates by two cells per iteration. The number of
iterations needed to converge to a solution is thus roughly
proportional to the dimension of the grid. On large grids this can
make the convergence of the algorithm very slow.

A standard technique to achieve faster convergence is to use {\it
  multigrid} acceleration. This means that a hierarchy of grids with
different resolutions is used. The basic concept is that the large
scale structure of the solution can be found quickly on a coarse grid,
so that the iterations on the finer grids only need to recover the
small scale properties which requires much fewer iterations. We adopt
this approach in our simulation code and use tree levels that are
coarser than the AMR level for the multigrid acceleration employing
the so-called {\it full approximation scheme} \citep{Brandt1977}.

In order to use this technique, we need an algorithm to map an
approximate solution $u$ to the next coarser or finer tree level. For
going to a coarser grid, also known as {\it restriction} and denoted
by the operator ${\cal R}$, we set the $u$ value of the parent node to
the average value of the daughter cells. To map to finer levels,
i.e. for the {\it prolongation} denoted by ${\cal P}$, we apply either
{\it straight injection}, which means that the $u$ value of the parent
node is assigned to all its daughter cells, or we compute the $u$
values using a linear reconstruction which is additionally based on
the gradient of $u$ in the centre of the parent cell.

The equation that needs to be solved for the multigrid acceleration on
the coarser grids differs slightly form
Eq.~(\ref{eq:u_discrete_field_eq}) \citep[see
  also][]{Oyaizu2008,Li2012}. Assume that we want to solve for $u$ on
a grid with grid spacing $h$. In the following, we make the grid
resolution explicit by using superscripts. We omit the cell indices
for convenience. Thus, Eq.~(\ref{eq:u_discrete_field_eq}) can be
rewritten as
\begin{equation}
  {\cal L}^h(u^h) = f^h. 
\label{eq:u_fine_grid}
\end{equation} 
Now assume that we have an approximate solution $\hat{u}^h$ which satisfies
\begin{equation}
  {\cal L}^h(\hat{u}^h) = \hat{f}^h.
\end{equation} 
Combining these two equations then yields
\begin{equation}
  {\cal L}^h(u^h) - {\cal L}^h(\hat{u}^h) = f^h - \hat{f}^h. 
\end{equation}
As a next step, we coarsen this equation in order to represent it on a
grid with two times larger grid spacing. We do this as follows,
\begin{equation}
  {\cal L}^{2h}(u^{2h}) = f^{2h},
\label{eq:u_coarse_grid}
\end{equation}
where
\begin{equation}
  f^{2h} = {\cal L}^{2h}({\cal R} \hat{u}^h) + {\cal R}(f^h - \hat{f}^h). 
\end{equation}
Eq.~(\ref{eq:u_coarse_grid}) has the same form as
Eq.~(\ref{eq:u_fine_grid}) and can thus be iteratively solved in a
similar fashion. For this our code uses ${\cal R} \hat{u}^h$ as an
initial guess. Large-scale features in the solution will then be
recovered much faster than on the fine grid. Once a better
approximation $\hat{u}^{2h}$ of $u^{2h}$ is found, the correction to
$u$ can be mapped back to the fine grid by
\begin{equation}
  \hat{u}^{h,{\rm new}} = \hat{u}^h + {\cal P}(\hat{u}^{2h}-{\cal R} \hat{u}^h).
\label{eq:fine_correction}
\end{equation}
Note that applying this correction to $\hat{u}^h$ rather than,
e.g. using $\hat{u}^{h,{\rm new}} = {\cal P}\hat{u}^{2h}$ has two
significant advantages. First, the small scale features in $\hat{u}^h$
are preserved. And second, the effects of an inaccurate mapping
between different levels are reduced. This can be most easily seen in
the case in which $\hat{u}^{2h}={\cal R} \hat{u}^h$, i.e. if the
coarse solution is not changed. It then directly follows from
Eq.~(\ref{eq:fine_correction}) that also the correction to the fine
solution is exactly equal to zero in spite of an imperfect mapping
between the two levels.

The individual steps the code takes for solving
Eq.~(\ref{eq:u_discrete_field_eq}) in a full timestep are as follows:
\newcounter{saveenum}
\begin{enumerate}[itemindent=7mm]
  \item It starts on the finest tree level that still covers the
    complete simulation volume. \label{item:first_step}
  \item The solution $u$ on this level from the previous timestep is
    used as an initial guess for the iterative solver.
  \item Using Eq.~(\ref{eq:iterative_correction}), the code performs
    $n_{\rm pre-smooth}$ red-black sweeps. \label{item:begin_V}
  \item The new $u$ values found in this way are mapped to the next
    coarser level. \label{item:map_coarse}
  \item The code solves the coarse-grid problem
    (Eq.~(\ref{eq:u_coarse_grid})). If the coarse-grid dimension is
    larger than $4^3$, this is again done using multigrid
    acceleration. Otherwise, the iterative Newton-Gauss-Seidel solver
    is directly employed.
  \item The change in the coarse grid solution is mapped back to the
    fine grid according to
    Eq.~(\ref{eq:fine_correction}). \label{item:map_fine}
  \item The code performs $n_{\rm post-smooth}$ red-black sweeps.
  \item If $|{\cal L}_{i,j,k} - f_{i,j,k}|$ is smaller than some
    threshold value $R_{\rm thres}$ for all $(i,j,k)$, $u$ is accepted
    as the solution for this tree level. Otherwise, steps
    \ref{item:begin_V} to \ref{item:end_level} are repeated,
    i.e. another so-called {\it V-cycle} is
    performed. \label{item:end_level}
    \setcounter{saveenum}{\value{enumi}}
\end{enumerate}

We have now found a solution $u$ on the finest level that covers the
complete simulation box. The code then proceeds to the next finer
level. The grid, there, typically consists of a number of patches and,
by definition, has boundaries as it does not cover the whole
volume. In the cells next to the boundary we define ${\cal L}_{i,j,k}$
using {\it ghost} cells. More precisely we add an additional layer of
such cells around each patch and compute their $u$ values by applying
the prolongation operation to their parent node. These ghost cells can
then be used as neighbours for the cells next to the boundary such
that ${\cal L}_{i,j,k}$ is defined for them and the iterative solver
can be applied. The ghost cells themselves are, however, not updated
during the iteration and retain the $u$ value that they were assigned
based on their parent node. More precisely, the solver works on levels
that do not cover the whole volume in the following way:

\begin{enumerate}[resume*]
\setcounter{enumi}{\value{saveenum}}
  \item The solution for $u$ previously found on the coarser level is
    mapped to the current tree level and used as an initial guess for
    the iterative solver, as well as for setting the $u$ values of the
    ghost cells. \label{item:start_fine}
  \item Eq.~(\ref{eq:u_discrete_field_eq}) on this level is then again
    solved using multigrid acceleration, i.e. performing steps
    \ref{item:begin_V} to \ref{item:end_level} for the current level.
  \item Once the solution has been found the code proceeds to the next
    finer level. \label{item:end_fine}
  \item Steps \ref{item:start_fine} to \ref{item:end_fine} are
    performed for each tree level down to the finest AMR
    level. \label{item:last_step}
\end{enumerate}

To summarize, our code solves Eq.~(\ref{eq:u_discrete_field_eq}) on
the finest tree level that still covers the complete simulation
volume, as well as on all finer levels that contain AMR cells. For
each level convergence can be accelerated using the multigrid
technique. The steps listed above give a good overview of how our code
works, however, some details still need to be filled in.

When we apply multigrid acceleration to grids that cover only part of
the simulation volume, we only want to iteratively update those
regions of the coarser grids that correspond to the same part of the
simulation box. We, thus, have to map the boundaries of the fine level
to the coarser levels. This is achieved by employing a mask function
(see, e.g. \citet{Guillet2011} for a detailed discussion on how to use
mask functions for this purpose). In our code, the mask function is
set to 1 in all cells on the level for which $u$ is currently
computed. The mask function on the coarser levels which are used for
the multigrid acceleration is then obtained by averaging the values of
the daughter cells. If a cell on the coarse grid does not have
daughter cells a mask value of 0 is assigned to it. We then include
all cells in the red-black sweeps that are performed for the multigrid
acceleration that have a mask value larger than some threshold value
$m_{\rm thres}$. We found $m_{\rm thres} = 0.4$ to work well and to
result in a fast convergence.

\subsection{Accounting for effective masses in the TreePM gravity}
\label{sec:treepm}

Once we have computed $f_R = \bar{f}_R(a) \times e^u$ for all AMR
cells, we can either compute the modified gravity accelerations
according to Eq.~(\ref{eq:modgrav_accel}) or include the corresponding
effective mass density in the Poisson equation for the gravitational
potential.

Adopting the latter approach, we first rewrite Eq.~(\ref{eq:phi_poisson_eq}) as
\begin{equation}
  \nabla^2 \phi = 4 \pi G \left(\delta \rho + \delta \rho_{\rm eff}
  \right),
\label{eq:phi_poisson_eq_with_eff_mass}
\end{equation}
where $\delta \rho_{\rm eff}$ is defined by
\begin{equation}
  \delta \rho_{\rm eff} = \frac{1}{3} \delta \rho - \frac{1}{24 \pi G}
  \delta R,
\label{eq:effective_density}
\end{equation}
and $\delta R$ is given by Eq.~(\ref{eq:dR_fR}). These equations imply
that the $f(R)$ modifications of gravity can be described as arising
from an ``effective mass density'' $\delta \rho_{\rm eff}$. We can,
thus, solve Eq.~(\ref{eq:phi_poisson_eq_with_eff_mass}) using the very
well tested and optimized TreePM-gravity algorithm that {\sc
  p-gadget3} uses for standard Newtonian gravity. The only modification that is
required is adding the effective mass density that we obtain from the
solution for $f_R$ to the standard matter density.

In a TreePM code the mass density enters the computation in two
places, that is in the masses assigned to the tree nodes and the
particle-mesh (PM) grid cells. From Eq.~(\ref{eq:effective_density})
we already know the additional effective mass terms for all tree nodes
that are AMR cells. The centre of effective mass is assumed to be the
cell centre in that case. For the parent nodes of AMR cells the
effective mass is computed by simply summing up the effective masses
of the daughter nodes. Their centres of effective mass are also
computed in a straightforward manner based on the effective masses and
centres of effective mass of the daughter cells. As effective masses
can be either positive or negative the centre of effective mass can in
rare cases fall outside the node. To avoid numerical inaccuracies due
to this, we always open such nodes in the tree walk.

The tree force computation is then done in the following way:
\begin{enumerate}[itemindent=7mm]

\item If a node on the AMR level or on coarser levels can be directly
  used\footnote{By which we mean that it does not need to be opened
    according to the geometrical or relative tree opening criterion.},
  the gravitational acceleration of the considered particle is updated
  including both the contributions from the real and effective masses
  of the tree node.

\item If an AMR node needs to be opened, the acceleration due to the
  AMR node's effective mass is directly applied, while the
  acceleration due to the real mass is computed based on the daughter
  nodes or on the particles contained in the tree node.

\end{enumerate}
The latter scheme accommodates the fact that the effective mass is not
defined below the AMR level.

For the long-range gravity, the effective mass of an AMR cell is
distributed among all overlapping PM grid cells. The fraction assigned
to an individual PM grid cell is simply given by the ratio of
overlapping to total AMR cell volume.

\section{$f(R)$-test problems}
\label{sec:tests}

Before applying our code to cosmological simulations, we want to make
sure it works reliably for simple test problems. To this end, we will
look at the $f_R$-fields of a point mass and a peaked
one-dimensional density distribution.

\subsection{The $f_R$-field of a point mass}
\label{sec:pointmass}

\begin{figure}
\centerline{\includegraphics[width=\linewidth]{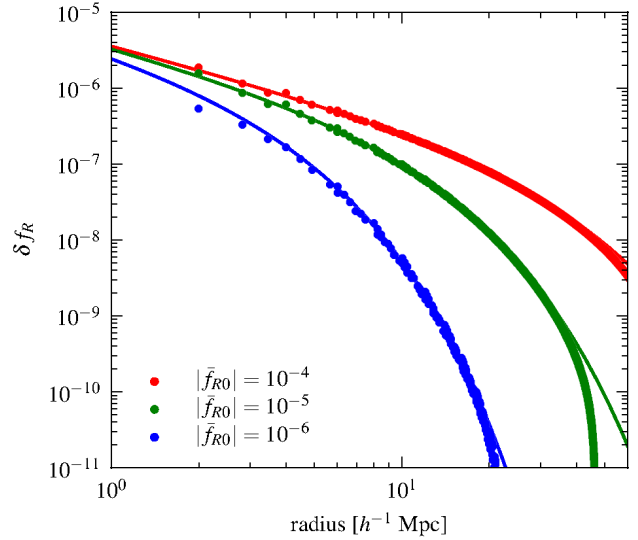}}
\caption{Solution for $\delta f_R$ found by our code ({\it circles})
  for a point mass for $|\bar{f}_R| = 10^{-4}$, $10^{-5}$ and a small
  spherical symmetric Gaussian mass distribution for $|\bar{f}_R| =
  10^{-6}$. The {\it solid} lines indicate the analytic solutions of
  the linearised equation for $\delta f_R$ assuming vacuum boundary
  conditions. They should be a good approximation of the full
  nonlinear solution with periodic boundary conditions in an
  intermediate radial range where nonlinearities and box size effects
  are not important. In that radial range we find excellent agreement
  with the numerical results.}
\label{fig:pointmass}
\end{figure}

For this test, we use essentially the same setup as in
\citet{Li2012}. More precisely, we use a $256 h^{-1} \textrm{Mpc}$ box
at $z=0$ covered by a uniform $N^3=128^3$ grid. The density in the
individual cells is given by
\begin{equation}
  \delta \rho = -10^{-4} \times \bar{\rho}
\end{equation}
for all cells except the cell where the point mass is placed. There
$\delta \rho$ is given by
\begin{equation}
  \delta \rho = 10^{-4} \times (N^3 - 1) \times \bar{\rho}.
\end{equation}
We then compute $f_R$ using our iterative solver on this uniform
grid. Figure~\ref{fig:pointmass} compares our numerical result to an
analytic solution, which is, however, only valid in an intermediate
radial range. The analytic calculation assumes that $\delta f_R = f_R
- \bar{f}_R$ is small compared to $|\bar{f}_R|$ so that
Eq.~(\ref{eq:dR_fR}) can be linearised. Then
Eq.~(\ref{eq:fR_field_eq}) reduces to a screened Poisson equation and
the solution for $\delta f_R$ is given by a Yukawa potential
\begin{equation}
  \delta f_R = \frac{2 G m}{3 c^2} \frac{e^{-\lambda r}}{r},
\end{equation}
where
\begin{equation}
  \lambda = \sqrt{\frac{1}{3 c^2} \frac{dR}{df_R} \bigg|_{f_R=\bar{f}_R}},
\end{equation}
and $m = 10^{-4} \times (N^3 - 1) \times \bar{\rho} \, V_{\rm cell}$
is the mass value of the point mass and $V_{\rm cell}$ is the volume
of a single cell. This solution is not valid very close to the point
mass where the assumed linearity breaks down. Furthermore it is also
inaccurate at very large radii, as there finite box size effects
become important. Note that the analytic derivation assumes vacuum
boundary conditions while the numerical solver adopts periodic
boundary conditions.

The figure displays results for different values of the background
value $\bar{f}_R$ of $f_R$. For $|\bar{f}_R| = 10^{-4}$ and $10^{-5}$
the assumed linearity still holds on the scale of the grid
spacing. Thus the numerical results are in very good agreement with
the analytic solution also at the smallest resolved radii. There is
only a small amount of scatter around the analytic solution at $r
\lesssim 8 \, \textrm{Mpc}/h$. This is, however, expected due to
discretisation errors as this region is only a few times larger than
the grid spacing. In an intermediate radial range there is almost
perfect agreement between the analytic and numerical solution. At
large radii, the solutions start to deviate from each other (most
noticeably in the $|\bar{f}_R| = 10^{-5}$ case) due to finite box size
effects. This is completely expected and we checked that the deviation
decreases when putting the same point mass in a larger box.

For $|\bar{f}_R| = 10^{-6}$, the analytic solution $\delta f_R$
exceeds $|\bar{f}_R|$ for $r \lesssim 2 \, \textrm{Mpc}/h$. This is
unphysical as $\delta R$ diverges for $f_R \rightarrow 0$. It is thus
obvious that the assumed linearity no longer holds at small
radii. Instead the Chameleon mechanism will start to screen $f_R$
there. This can however not be resolved numerically for a point mass
that resides only in a single cell. Therefore, for the
$|\bar{f}_R| = 10^{-6}$ case we have replaced the point mass by a small
spherically symmetric Gaussian mass distribution $\delta \rho \sim
\exp[-r^2/(2 \, \textrm{Mpc}/h)^2]$ with the same total mass. Then the
numerical result is in good agreement with the analytic solution for a
true point mass except for the expected deviations at very small and
very large radii.

Overall, these tests show that our numerical solver yields results
that are in excellent agreement with the analytic solutions in the
radial ranges in which the latter are valid. The deviations at very
small and large radii are completely expected and are caused by
assuming linearity and vacuum boundary conditions in the analytic
derivation.

\subsection{The $f_R$-field of a 1D density peak}
\label{sec:gauss1D}

With the next test problem we want to check whether our solver also
works reliably when using a mesh that adaptively refines on density
peaks. To this end, we set up a 1D density peak in a $256 \, h^{-1}
\textrm{Mpc}$ box. More precisely we use a density field that depends
only on the $x$-coordinate and has a peak at $x = 128 \, h^{-1}
\textrm{Mpc}$. The shape of the peak is chosen such that it
corresponds to the following $f_R$ field
\begin{equation}
  f_R = -A \times \left(1 - \alpha e^{-\frac{\Delta x^2}{w^2}}\right),
\label{eq:fR_gauss_1D}
\end{equation}
where $\Delta x$ is the distance from the density peak and $w=15 \,
h^{-1} \textrm{Mpc}$. $\alpha = 0.99999$, i.e. close to unity, was
chosen to test our code deep in the nonlinear regime. We adopted
$|\bar{f}_R| = 10^{-5}$ for this test. The corresponding density
perturbation is then given by
\begin{equation}
  \delta \rho = \frac{1}{8 \pi G} \left( \delta R(f_R) - 3 c^2
  \frac{{\rm d}^2f_R}{{\rm d}x^2} \right)
\end{equation}
according to Eq.~(\ref{eq:fR_field_eq}), where $\delta R(f_R)$ is
given by Eq.~(\ref{eq:dR_fR}). In this way it is easy to find an
analytic expression for $\delta \rho$. We then choose $A \approx 2.7
\times 10^{-5}$ such that the spatial average of $\delta \rho$
vanishes. As a next step, we sample this density field with an
adaptive grid with $256^3$ base grid cells and two levels of
refinement. This enables us to use it as an input to our
adaptive-resolution multigrid-accelerated solver.

\begin{figure}
\centerline{\includegraphics[width=\linewidth]{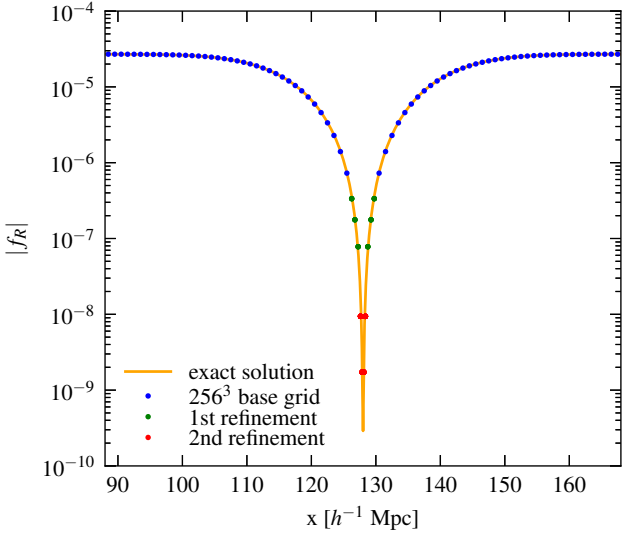}}
\caption{Solution for $\delta f_R$ found by our AMR code ({\it
    circles}) for a 1D density peak and $|\bar{f}_R| = 10^{-5}$. The
  analytic solution is indicated by the {\it solid} line. The colours
  of the {\it circles} encode the refinement level of the individual
  AMR cells. The numerical solution on all AMR levels is in excellent
  agreement with the analytic prediction.}
\label{fig:gauss1D}
\end{figure}

In Fig.~\ref{fig:gauss1D}, we display the solution which it finds for
$f_R$ and compare it to the analytic solution given by
Eq.~(\ref{eq:fR_gauss_1D}). They are in excellent agreement on the
base grid, as well as on the refined grids. This demonstrates that our
relaxation solver and the AMR algorithm work reliably for such a
simple test problem. Further tests of our AMR code for the more
realistic case of a cosmological density field are presented in
Sec.~\ref{sec:AMR_test}.

\section{Cosmological f(R) N-body simulations}
\label{sec:cosmo_runs}

In the \citet{Hu2007} model modifications of gravity appear only at
low redshift. This can be understood from Eq.~(\ref{eq:fR_mean}). At
early times, i.e. for $a \ll 1$, $\bar{f}_R(a)$ is very small which
essentially restores standard $\Lambda$CDM. We can, thus, start our
cosmological $f(R)$-simulations from the same initial conditions that we
use for $\Lambda$CDM runs. In the following we present further code
tests and first results from full cosmological simulations performed
with {\sc mg-gadget}.

\subsection{Testing adaptive mesh refinement for a cosmological density field}
\label{sec:AMR_test}

In Sec.~\ref{sec:gauss1D}, we have tested our AMR algorithm by
comparing its results to a known analytic solution. In more realistic
situations in which an analytic solution is not known, it is in
principle possible to validate the AMR method by comparison to results
obtained on a very-high resolution fixed grid. In practice, however,
obtaining a solution on a fixed grid with a resolution equal to the
peak resolution of an AMR code may be computationally prohibitively
expensive. To circumvent this problem, we perform such a comparison
only for a single timestep and additionally limit the maximum
refinement of the AMR code.

\begin{figure}
\centerline{\includegraphics[width=\linewidth]{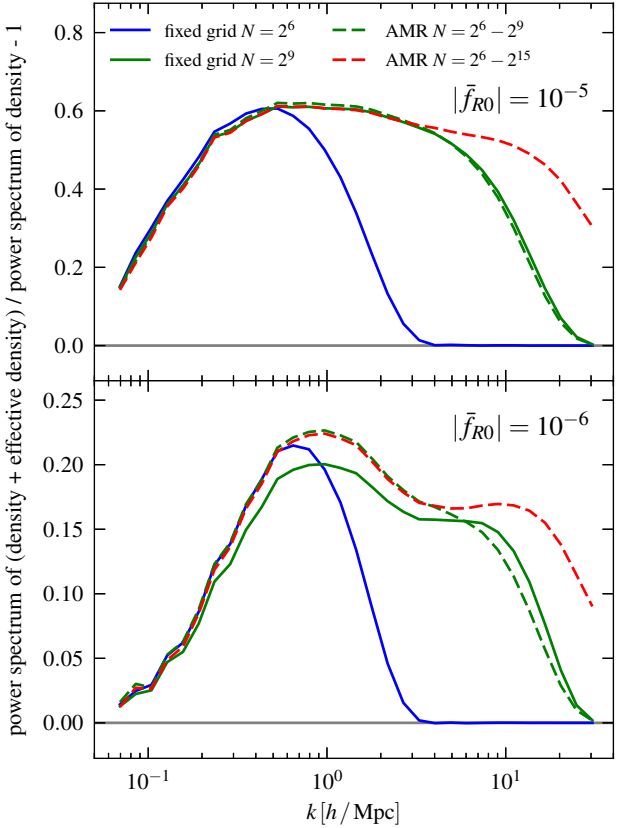}}
\caption{Enhancement of gravity as a function of scale. More
  precisely, we show the ratio of the power spectrum of $\delta \rho +
  \delta \rho_{\rm eff}$ to the power spectrum of $\delta \rho$ for
  $|\bar{f}_{R0}|=10^{-5}$ ({\it upper panel}) and $10^{-6}$ ({\it
    lower panel}) for a cosmological density field at $z=0.5$. The
  effective density was computed for a co-moving $100 h^{-1}
  \textrm{Mpc}$ box either on a fixed grid or using the AMR
  technique. For the fixed grid results the grid resolution is quoted
  in the figure's legend. For the AMR runs the base and peak
  resolutions are quoted. At the same peak resolution there is
  excellent agreement between high-resolution fixed grid and AMR runs
  for $|\bar{f}_{R0}|=10^{-5}$. For $|\bar{f}_{R0}|=10^{-6}$ small
  deviations are visible which are due to fundamental limitations of
  the accuracy of {\it one-way interface} scheme AMR methods for very
  nonlinear problems. These discrepancies can, however, be alleviated
  by choosing a higher base resolution.}
\label{fig:eff_density_pow_sepc}
\end{figure}

In Fig.~\ref{fig:eff_density_pow_sepc}, we compare the power spectrum
of $\delta \rho + \delta \rho_{\rm eff}$, where $\delta \rho_{\rm
  eff}$ is defined by Eq.~(\ref{eq:effective_density}), to the power
spectrum of $\delta \rho$ alone. The difference between the two can be
interpreted as a measure of the enhancement of gravity as a function
of scale. The comparison is based on a density field obtained from a
full cosmological simulation at $z=0.5$. $f_R$ and $\delta \rho_{\rm
  eff}$ are computed for models which, at $z=0$, have a background
value of $|\bar{f}_{R0}|=10^{-5}$ and $10^{-6}$. $f_R$ is found either
on a low-resolution $64^3$ or high-resolution $512^3$ fixed grid or
using the AMR method with a $64^3$ base grid and a peak resolution
equalling that of a $512^3$ and a $16384^3$ fixed grid. The refinement
in the former AMR approach is thus restricted to yield the same peak
resolution as the high-resolution fixed grid, while in the latter case
the full AMR algorithm is employed.

For $|\bar{f}_{R0}|=10^{-5}$, the restricted AMR computation is in
almost prefect agreement with the results from the high-resolution
fixed grid. This demonstrates that our AMR algorithm works well and
recovers in a computationally much more efficient way the solution
that one would obtain with a high-resolution fixed grid. As expected,
using the full AMR algorithm results in an enhancement of gravity on
even smaller spatial scales that can not be resolved by the
high-resolution fixed grid. It is also worth noting that the low- and
high-resolution fixed grid results are in excellent agreement with
each other on scales $k<0.5 \, h / {\rm Mpc}$, which are well resolved
on both grids.

For $|\bar{f}_{R0}|=10^{-6}$, the restricted AMR computation is still
in reasonably good agreement (better than $\sim12\%$ accuracy in the
modification of gravity on well resolved scales $k\lesssim 5 \, h / {\rm
  Mpc}$) with the results obtained on the high-resolution fixed
grid. There are, however, also noticeable deviations.

Interestingly, deviations of similar magnitude exist between the low-
and high-resolution fixed grid calculations on large scales $k<0.5 \,
h / {\rm Mpc}$, which are in principle well resolved on both
grids. This can be understood in the following way. For
$|\bar{f}_{R0}|=10^{-5}$, the source term on the right-hand side of
Eq.~(\ref{eq:fR_field_eq}) is still rather linear in $\delta f_R$, so
that coarse-graining Eq.~(\ref{eq:fR_field_eq}) to the low-resolution
grid does not significantly change the solution on large scales. For
$|\bar{f}_{R0}|=10^{-6}$, the source term is much more nonlinear which
results in a slightly different coarse-grained solution. Physically,
this most likely means that some of the objects in which the chameleon
effect occurs are not properly resolved on the low-resolution grid,
which somewhat changes the solution.

Like most other multigrid-accelerated AMR gravity solvers (including
the {\sc ramses} solver \citep{Guillet2011} on which the code
presented in \citet{Li2012} is based), our AMR algorithm uses a {\it
  one-way interface} scheme. This means that the solution on a coarse
grid is used as a boundary condition for the refined patches on the
next finer grid level, while the solution on the fine grid does not
react back on the coarse grid solution in unrefined regions. In
particular, the solution that the AMR solver yields in regions that
are only covered by the base grid is identical to the one which would
be obtained on a fixed grid with the same resolution as the base
grid. This is also the reason why the AMR runs yield the same results
as the low-resolution fixed grid calculations on the largest
scales. There, the deviations between the low- and high-resolution
fixed grid runs for $|\bar{f}_{R0}|=10^{-6}$, thus, necessarily result
in deviations between the restricted AMR run and the high-resolution
fixed grid results. This in turn means that there are fundamental
limitations on the accuracy of {\it one-way interface} AMR schemes for
very nonlinear equations.

Note, however, that for this test, we have chosen a rather low base
resolution ($\sim 1.5 \, h^{-1}$ Mpc co-moving grid spacing) for the
AMR runs. Using a higher base resolution, as done throughout the
remainder of this paper alleviates this problem. Nevertheless, in the
future, it may be worthwhile to develop a code which uses a {\it
  two-way interface} AMR scheme in which the fine grid solution acts
back onto the coarse grid solution, e.g. by matching both $f_R$ and
its derivative at the coarse/fine-boundary, for strongly nonlinear
models. For {\it one-way interface} AMR schemes we recommend using a
relatively high base resolution to increase the accuracy for such
models.  Overall, our AMR scheme works extremely well for mildly
nonlinear models and still very well for $|\bar{f}_{R0}|=10^{-6}$, in
particular when using a sufficiently fine base grid.

\subsection{The matter power spectrum}
\label{sec:matter_power_spec}

The matter power spectrum is a sensitive and widely used statistics to
describe structure formation in the Universe. In the context of
modified gravity models it is, thus, one of the first quantities to
investigate.

\begin{figure}
\centerline{\includegraphics[width=\linewidth]{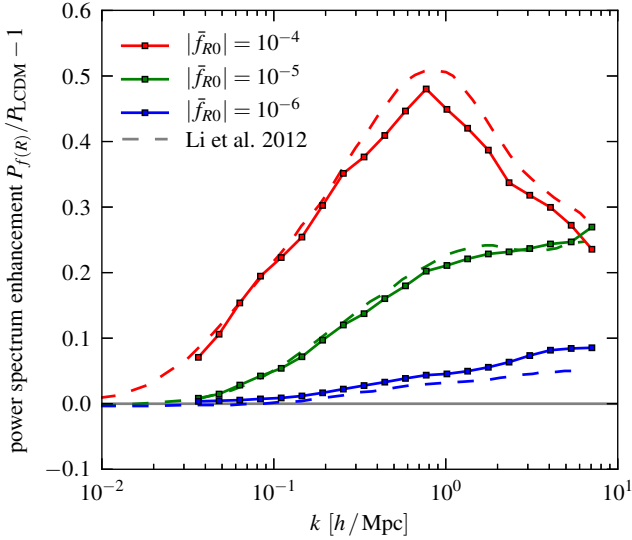}}
\caption{Matter power spectrum enhancements with respect to a standard
  $\Lambda$CDM model at $z=0$ for $|\bar{f}_{R0}|=10^{-4}$, $10^{-5}$
  and $10^{-6}$ ({\it solid lines} and {\it squares}). Results
  obtained in $1000 \, h^{-1}$ Mpc boxes by \citet{Li2012} are shown
  for reference ({\it dashed lines}).}
\label{fig:pow_enhance}
\end{figure}

In Fig.~\ref{fig:pow_enhance}, we display the enhancement of the
matter power spectrum in the \citet{Hu2007} model for $n=1$ and
$|\bar{f}_{R0}|=10^{-4}$, $10^{-5}$ and $10^{-6}$, compared to a
standard $\Lambda$CDM universe. All runs where started from identical
initial conditions at $z=100$. The initial conditions are consistent
with a WMAP-7yr cosmology \citep{Komatsu2011}, i.e. $\Omega_{\rm
  M}=0.272$, $\Omega_{\Lambda} = 0.728$, $\Omega_{\rm B}=0.0456$,
$h=0.704$, and $\sigma_{8}=0.809$. A co-moving box size of $200 \,
h^{-1}$ Mpc, $256^3$ particles, the same number of base grid cells and
up to eight levels of refinement were used. Our results are compared
to \citet{Li2012}.

In general, the power spectrum enhancements we find are in
impressively good agreement with those obtained by
\citet{Li2012}. This is quite reassuring. Minor deviations that do
exist on small scales are not completely unexpected. First, we would
like to point out that in this comparison cosmic variance does not
appear on the largest scales but on intermediate and small scales. The
reason for this is that the $\Lambda$CDM and modified gravity runs
were started from identical initial conditions. Thus, on large scales
in the linear regime, where mode-coupling is negligible, there is no
cosmic variance as the initial phases and amplitudes are identical. On
smaller scales, where mode-coupling becomes important, cosmic variance
can appear because the enhancement of the amplitude of a particular
mode depends on the amplitudes and phases of other modes. For our box
size of only $200 \, h^{-1}{\rm Mpc}$, we do, thus, expect some cosmic
variance on intermediate and small scales. In addition, there may be a
mild dependence on the resolution of the simulation as suggested by
the discussion in Sec.~\ref{sec:AMR_test}. These deviations and
effects are in agreement with the slightly different power spectrum
enhancements obtained on small scales for simulations with different
box sizes and resolutions in \citet{Li2012} (see their Fig.~12).
Overall, it is very reassuring that the power spectrum enhancements
obtained with {\sc mg-gadget} are so strikingly similar to those found
by \citet{Li2012}.

\section{Cosmological hydrodynamical f(R) simulations}
\label{sec:cosmo_hydro_runs}

So far almost all simulations of modified gravity models ignored the
effects of baryonic physics. Here, we present the first -- to our knowledge --
non-radiative and radiative cosmological hydrodynamical
simulations with modified gravity. In the following we discuss why
such a simultaneous treatment of modified gravity and baryonic physics
will be instrumental for successfully probing modifications of gravity
with upcoming surveys like Euclid or LSST.

\subsection{Degeneracies between baryonic physics and modifications of gravity}
\label{sec:baryon_vs_modgrav}

\begin{figure}
\centerline{\includegraphics[width=\linewidth]{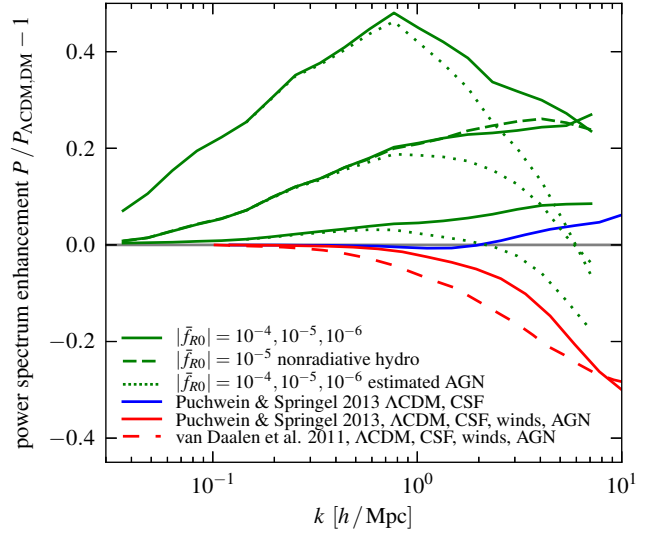}}
\caption{Comparison of the effects of modified gravity and baryonic
  physics on the total matter power spectrum at $z=0$. All power
  spectrum enhancements are with respect to a collisionless
  $\Lambda$CDM simulation. AGN feedback affects the total matter power
  spectrum on similar scales as modified gravity. The magnitudes of
  the effects are also comparable. This illustrates that there are
  significant degeneracies between uncertainties in the baryonic
  physics and the impacts of modified gravity.}
\label{fig:baryonic_physics}
\end{figure}

In Fig.~\ref{fig:baryonic_physics}, we compare the effects of modified
gravity and baryonic physics on the total matter power spectrum. The
{\it green solid} lines show the same power spectrum enhancements due
to modified gravity as in Fig.~\ref{fig:pow_enhance}, which were
computed from collisionless simulations. The {\it green dashed} line
displays the enhancement we obtain for a non-radiative hydrodynamical
modified-gravity run of the same $200 \, h^{-1}{\rm Mpc}$ box with
$|\bar{f}_{R0}|=10^{-5}$. The hydrodynamics was followed with {\sc
  p-gadget3}'s entropy-conserving formulation of smoothed particle
hydrodynamics \citep{Springel2002}. As expected, the power spectrum
enhancement in this non-radiative hydrodynamical simulation is very
similar to the one obtained in the collisionless run with the same
$f_{R0}$ value.

The {\it blue solid} line shows the effects of radiative cooling and
star formation, which are treated using the \citet{Springel2003}
model, on the total matter power spectrum in a $\Lambda$CDM
universe. The curve was obtained by comparing simulation {\it S13}
from \citet{Puchwein2013} to a collisionless simulation of the same
box. In this run there is an increase of the total matter power
spectrum on small scales due to the radiative collapse of baryons in
halo centres, as well as the consequently triggered adiabatic contraction
of the dark matter distribution. Note, however, that this simulation
lacks an efficient feedback mechanism and therefore suffers from
overcooling.

The {\it red solid} line shows the effects of baryonic physics in
$\Lambda$CDM simulations that in addition to radiative cooling and
star formation account for supernovae-driven winds and energy
injection by supermassive black holes, i.e. for active galactic nuclei
(AGN) feedback which is treated essentially as in
\citet{Sijacki2007}\footnote{See \citet{Puchwein2013} for details on
  minor modifications with respect to the AGN feedback model in
  \citet{Sijacki2007}}. This model, which accounts for both quasar- and
radio-mode AGN feedback, was shown to result in a self-regulated black
hole growth \citep{DiMatteo2005,Sijacki2007} and results in much more
realistic properties of galaxy clusters \citep{Puchwein2008} and their
central galaxies \citep{Puchwein2010}, as well as abundances of
massive galaxies \citep{Puchwein2013}. The curve is based on a
simulation with essentially the same parameters as the {\it S16} run
from \citet{Puchwein2013}. The only difference is that a larger seed
black hole mass of $10^5 h^{-1} M_\odot$ was used, which results in a
slightly better agreement of cluster and group properties with
observations. As the power spectrum at Mpc scales is dominated by the
one-halo term of these objects \citep[e.g.][]{Semboloni2011}, this
seems to be a better choice for the comparison presented here. It can
clearly be seen that at least in this model AGN feedback significantly
suppresses the total matter power spectrum on scales smaller than
$\sim 6 \, h^{-1}$ Mpc. This qualitatively confirms a similar finding
by \citet{VanDaalen2011}. Their result is shown for comparison ({\it
  dashed red} line).

Given that the AGN feedback prescription in \citet{VanDaalen2011} and
\citet{Puchwein2013} are partly based on the same assumptions, i.e. an
accretion model motivated by Bondi-Hoyle-Lyttleton accretion and
thermal injection of the feedback energy, the difference between these
results should probably be considered as a lower limit of the
theoretical uncertainty of the impact of AGN feedback. Understanding
these baryonic effects on the total matter power spectrum and their
uncertainties will be of utmost importance for fully exploiting
surveys like Euclid or LSST that aim to probe cosmology by weak
lensing \citep[e.g.][]{Semboloni2011}. These results suggest that,
ultimately, the smallest spatial scales that can be used for such
studies may be determined by our (limited) understanding of baryonic
physics.

If, for the moment, we boldly assume that the relative suppression of
the total matter power spectrum by baryonic physics, most notably by
AGN feedback, is the same in modified gravity as in the $\Lambda$CDM
simulations from \citet{Puchwein2013}, we obtain the {\it green
  dotted} curves which show the expected power in modified gravity
models when accounting for all the aforementioned baryonic
physics. These results illustrate that baryonic physics and modified
gravity affect the matter distribution on similar scales. Especially
for the $|\bar{f}_{R0}|=10^{-6}$ model, also the magnitudes of the
effects are very similar. Furthermore, the $|\bar{f}_{R0}|=10^{-4}$
and $10^{-5}$ models are already in tension with observational
constraints on the scales of galaxies and smaller (see, e.g. Fig.~9 in
\citet{Lombriser2012} for an overview of observational constraints).

This comparison, therefore, shows that there are significant
degeneracies between the effects of uncertain baryonic physics and
modifications of gravity for models that are fully consistent with the
data. Simulations that include both effects will, hence, be required
for studying these degeneracies and for looking for signatures of
modified gravity that are least affected by uncertainties in the
baryonic physics.

The current version of {\sc mg-gadget} does in principle allow
following the dynamical evolution in modified gravity models and
processes like AGN feedback at the same time. However, such
simulations require large volumes for covering all the scales in which
we are interested and for reducing cosmic variance, as well as high
resolution, so that the sub-resolution feedback model yields converged
results. We, therefore, plan to put some more effort in further
improving the code performance in situations with a high dynamic range
before embarking on such an endeavour. As a first step in this
direction, we have performed radiative cosmological hydrodynamical
simulations with modified gravity and a strongly simplified treatment
of star formation. We present them below.

\subsection{Radiative hydrodynamical simulations with f(R)-gravity}
\label{sec:Lya}

Using {\sc mg-gadget} we performed the first radiative cosmological
hydrodynamical simulations with $f(R)$-gravity. Radiative cooling and
photoheating was included under the assumptions of a primordial gas
decomposition and of ionisation equilibrium in the presence of an
external UV background field as in \citet{Katz1996}. Star formation was
followed with a strongly simplified treatment \citep{Viel2004} in
which all gas particles exceeding a density of 1000 times the mean
baryon density and having a temperature lower than $10^5$ K are
transformed into collisionless star particles. While this scheme may
not produce realistic galaxy populations, it does allow obtaining
reliable predictions for the intergalactic medium (IGM). For these
simulations a co-moving box size of $40 \, h^{-1}$ Mpc, an initial
particle number of $2 \times 256^3$ (half of them gas and half of them
dark matter), $256^3$ base grid cells and up to eight levels of
refinement were used.

\begin{figure*}
\centerline{\includegraphics[width=\linewidth]{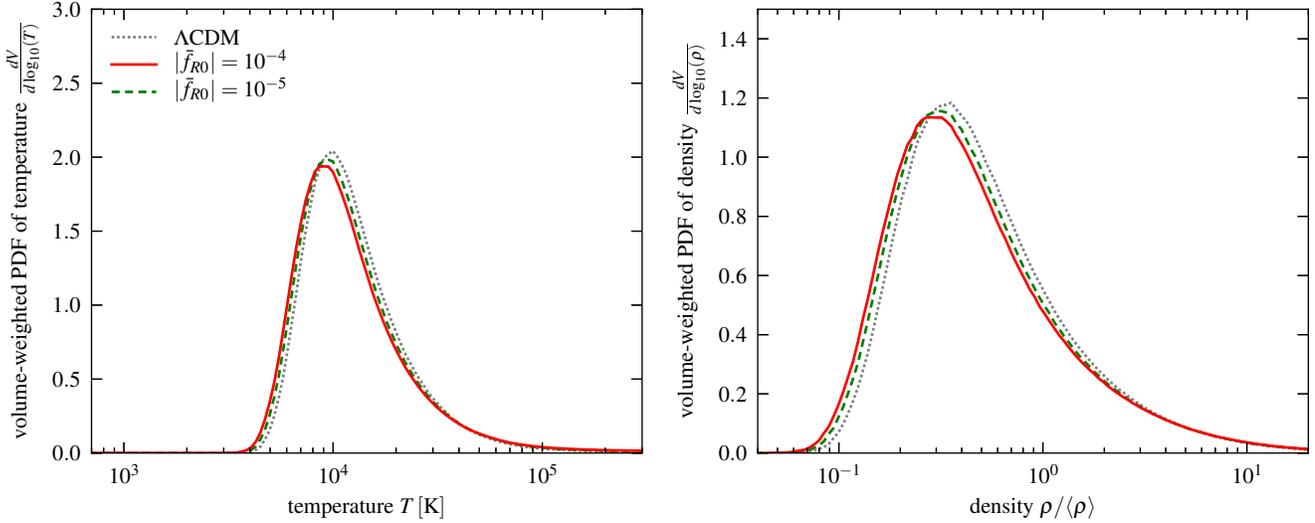}}
\caption{Volume-weighted PDF of gas temperature ({\it left panel}) and
  density ({\it right panel}) of the intergalactic medium in
  simulations with a standard $\Lambda$CDM cosmology and for modified
  gravity runs with $|\bar{f}_{R0}|=10^{-4}$ and $10^{-5}$ at redshift
  $z=2$.}
\label{fig:T_rho_pdfs}
\end{figure*}

Fig.~\ref{fig:T_rho_pdfs} shows the volume-weighted probability
distribution functions (PDFs) of the temperature and density of the
intergalactic medium at $z=2$. Differences between the $\Lambda$CDM
and the modified gravity runs are visible in both PDFs. The effects on
the temperature PDF, e.g. the $~15\%$ shift in the peak temperature,
may, however, be rather small compared to uncertainties in the heating
efficiency due to helium {\sc II} reionization \citep[see
  e.g.][]{McQuinn2009} or the effects of a recently suggested
additional heating of the IGM by TeV-blazars
\citep{Broderick2012,Puchwein2012}.

In contrast, the effect of modified gravity on the gas density PDF is
most likely much less degenerate with baryonic processes. The
enhancement of gravity in the $f(R)$ runs accelerates structure
formation and results in voids that are emptier than in $\Lambda$CDM,
as well as in a larger fraction of dense regions, although the latter
is somewhat hard to see in the figure.

These effects on the IGM density distribution may have observable
signatures in the Lyman-$\alpha$ forest. For example, the enhancement
of the matter power spectrum may be reflected by a change in the
Lyman-$\alpha$ forest transmitted flux power spectrum. It will be
interesting to investigate this in more detail in the future.

\section{Summary and conclusions}
\label{sec:conclusions}

We have presented the novel simulation code {\sc mg-gadget} which is
well suited for performing large cosmological N-body and
hydrodynamical simulations of modified gravity models. The code is
massively parallel and uses a multigrid-accelerated relaxation scheme
on an adaptively-refined mesh to solve for perturbations in the scalar
degree of freedom of the considered modified gravity model. The code
is implemented as a module for the widely used cosmological simulation
code {\sc p-gadget3}. This allows us to account at the same time for
baryonic processes like hydrodynamics or radiative cooling that
{\sc p-gadget3} is able to follow.

As a first application we consider the chameleon-type $f(R)$-gravity
model introduced by \citet{Hu2007}. Before performing full
cosmological runs, we apply the code to a few simple test problems to
assess the accuracy of the modified gravity solver. For a single point
mass, the numerical results are in very good agreement with the
analytic prediction in the radial range in which the latter is
valid. The code also accurately recovers the analytic solution for the
scalar degree of freedom in the case of a 1D density peak. For this
test, we use adaptive refinement. The numerical and analytic solutions
are in excellent agreement both on the base grid, as well as on the
refined grids, thereby demonstrating that our AMR scheme works well.

We also test the AMR method under more realistic conditions, i.e. for
a cosmological density field. Overall, it performs very well. For
$|\bar{f}_{R0}|=10^{-5}$, it yields essentially the same result as
obtained on an extremely fine fixed grid with a resolution equalling
the peak resolution of the adaptive mesh. However, this test also
reveals fundamental limitations of the accuracy of {\it one-way}
interface scheme AMR methods for very nonlinear equations. For the
considered $f(R)$-gravity model strong nonlinearities occur for small
$|\bar{f}_{R0}|$ values. For $|\bar{f}_{R0}|=10^{-6}$, small
deviations between the extremely fine fixed grid and AMR solutions
become noticeable. This discrepancy can, however, be alleviated by
using a higher base grid resolution.

As a next step, we perform full cosmological $f(R)$-gravity N-body
simulations and compare them to a reference $\Lambda$CDM run. We find
scale-dependent enhancements of the matter power spectrum due to the
modifications of gravity which are in very good agreement with results
obtained by \citet{Li2012}. This concordance of different
modified-gravity simulation codes is quite reassuring.

We then confront the effects of baryonic processes, like AGN feedback,
on the total matter power spectrum with its changes due to
modifications of gravity.  It turns out that for those modified
gravity models that are not in tension with observational constraints,
both effects have similar magnitude and happen at similar spatial
scales. This clearly demonstrates that there are significant
degeneracies between modified gravity and uncertainties in the
baryonic physics. We hence require simulations that follow both
processes at the same time to look for observational signatures of
modified gravity that are least affected by such uncertainties.

For the aforementioned comparison, we have analysed the effects of AGN
feedback on the total matter power spectrum in cosmological
hydrodynamical $\Lambda$CDM simulations by \citet{Puchwein2013}. Our
results confirm the finding of \citet{VanDaalen2011} that AGN feedback
significantly suppresses the total matter power spectrum on scales up
to several Mpc. Understanding and accounting for this will be of
utmost importance for fully exploiting upcoming surveys like Euclid or
LSST that aim to probe cosmology by weak lensing
\citep[e.g.][]{Semboloni2011}. Given these results, it seems likely
that the smallest spatial scale that can be used for such studies will
ultimately be set by our knowledge or lack of knowledge of baryonic
processes like feedback from AGN.

Finally, as a step towards simulations that account for modified
gravity and complex baryonic physics at the same time, we perform the first
-- to our knowledge -- modified gravity cosmological hydrodynamical
simulations. Some of our runs also account for radiative cooling in
the presence of an external UV background. In the latter runs, we find
that modified gravity changes the density and temperature PDFs of the
intergalactic medium. It will be interesting to explore in a future
work whether this results in an observable signature of modified
gravity in the Lyman-$\alpha$ forest.

\section*{Acknowledgements}

We are grateful to Luca Amendola and Jochen Weller for illuminating discussions. E.P. would also like to thank R\"udiger Pakmor, Andreas Bauer and Christoph Pfrommer for many helpful conversations. E.P. acknowledges support by the DFG through Transregio 33. M.B. is supported by the Marie Curie Intra European Fellowship ``SIDUN"
within the 7th Framework Programme of the European Community. MB also
acknowledges partial support  by the DFG Cluster of Excellence ``Origin and
Structure of the Universe'' and by the TRR33 Transregio Collaborative
Research Network on the ``Dark Universe''. 
V.S. acknowledges support through SFB 881, `The Milky Way System',
of the DFG.

\appendix

\bibliographystyle{mn2efixed}
\bibliography{paper}

\begin{thebibliography}{86}
\expandafter\ifx\csname natexlab\endcsname\relax\def\natexlab#1{#1}\fi

\bibitem[{Abbott {et~al}\mbox{.}(2005)Abbott {et~al.}}]{DES}
Abbott T., {et~al.}, 2005, arXiv:astro-ph/0510346

\bibitem[{Ade {et~al}\mbox{.}(2013)Ade {et~al.}}]{Planck_016}
Ade P., {et~al.}, 2013, arXiv:1303.5076

\bibitem[{Amendola(2000)}]{Amendola_2000}
Amendola L., 2000, Phys. Rev. D, 62, 043511

\bibitem[{{Amendola} {et~al}\mbox{.}(2012){Amendola}, {Appleby}, {Bacon},
  {Baker}, {Baldi}, {Bartolo}, {Blanchard}, {Bonvin}, {Borgani}, {Branchini},
  {Burrage}, {Camera}, {Carbone}, {Casarini}, {Cropper}, {deRham}, {di Porto},
  {Ealet}, {Ferreira}, {Finelli}, {Garcia-Bellido}, {Giannantonio}, {Guzzo},
  {Heavens}, {Heisenberg}, {Heymans}, {Hoekstra}, {Hollenstein}, {Holmes},
  {Horst}, {Jahnke}, {Kitching}, {Koivisto}, {Kunz}, {La Vacca}, {March},
  {Majerotto}, {Markovic}, {Marsh}, {Marulli}, {Massey}, {Mellier}, {Mota},
  {Nunes}, {Percival}, {Pettorino}, {Porciani}, {Quercellini}, {Read},
  {Rinaldi}, {Sapone}, {Scaramella}, {Skordis}, {Simpson}, {Taylor}, {Thomas},
  {Trotta}, {Verde}, {Vernizzi}, {Vollmer}, {Wang}, {Weller}, \&
  {Zlosnik}}]{Euclid_TWG}
{Amendola} L. {et~al.}, 2012, arXiv:1206.1225

\bibitem[{{Amendola} {et~al}\mbox{.}(2008){Amendola}, {Baldi}, \&
  {Wetterich}}]{Amendola_Baldi_Wetterich_2008}
{Amendola} L., {Baldi} M., {Wetterich} C., 2008, \prd, 78, 023015

\bibitem[{Armendariz-Picon {et~al}\mbox{.}(2001)Armendariz-Picon, Mukhanov, \&
  Steinhardt}]{kessence}
Armendariz-Picon C., Mukhanov V.~F., Steinhardt P.~J., 2001, Phys. Rev. D, 63,
  103510

\bibitem[{{Baldi}(2011)}]{Baldi_2011a}
{Baldi} M., 2011, \mnras, 411, 1077

\bibitem[{Baldi(2012)}]{Baldi_2012b}
Baldi M., 2012, Phys. Dark Univ., 1, 162

\bibitem[{{Barnes} \& {Hut}(1986)}]{Barnes1986}
{Barnes} J., {Hut} P., 1986, \nat, 324, 446

\bibitem[{Batista \& Pace(2013)}]{Batista_Pace_2013}
Batista R., Pace F., 2013, arXiv:1303.0414

\bibitem[{Baumann {et~al}\mbox{.}(2012)Baumann, Nicolis, Senatore, \&
  Zaldarriaga}]{Baumann_etal_2012}
Baumann D., Nicolis A., Senatore L., Zaldarriaga M., 2012, JCAP, 1207, 051

\bibitem[{Behrend {et~al}\mbox{.}(2008)Behrend, Brown, \&
  Robbers}]{Behrend_Brown_Robbers_2008}
Behrend J., Brown I.~A., Robbers G., 2008, JCAP, 0801, 013

\bibitem[{Bertotti {et~al}\mbox{.}(2003)Bertotti, Iess, \&
  Tortora}]{Bertotti_Iess_Tortora_2003}
Bertotti B., Iess L., Tortora P., 2003, Nature, 425, 374

\bibitem[{Brandt(1977)}]{Brandt1977}
Brandt A., 1977, Mathematics of Computation, 31, 333

\bibitem[{Brax {et~al}\mbox{.}(2011)Brax, van~de Bruck, Davis, Li, \&
  Shaw}]{Brax_etal_2011}
Brax P., van~de Bruck C., Davis A.-C., Li B., Shaw D.~J., 2011, Phys. Rev. D,
  83, 104026

\bibitem[{{Broderick} {et~al}\mbox{.}(2012){Broderick}, {Chang}, \&
  {Pfrommer}}]{Broderick2012}
{Broderick} A.~E., {Chang} P., {Pfrommer} C., 2012, \apj, 752, 22

\bibitem[{{Buchdahl}(1970)}]{Buchdahl_1970}
{Buchdahl} H.~A., 1970, \mnras, 150, 1

\bibitem[{Caldwell(2002)}]{Caldwell_2002}
Caldwell R., 2002, Physics Letters B, 545, 23

\bibitem[{{Casarini} {et~al}\mbox{.}(2011){Casarini}, {Macci{\`o}},
  {Bonometto}, \& {Stinson}}]{Casarini_etal_2011}
{Casarini} L., {Macci{\`o}} A.~V., {Bonometto} S.~A., {Stinson} G.~S., 2011,
  \mnras, 412, 911

\bibitem[{Clarkson {et~al}\mbox{.}(2011)Clarkson, Ellis, Larena, \&
  Umeh}]{Clarkson_etal_2011}
Clarkson C., Ellis G., Larena J., Umeh O., 2011, Rept. Prog. Phys., 74, 112901

\bibitem[{Creminelli {et~al}\mbox{.}(2009)Creminelli, D'Amico, Norena, \&
  Vernizzi}]{Creminelli_etal_2009}
Creminelli P., D'Amico G., Norena J., Vernizzi F., 2009, JCAP, 0902, 018

\bibitem[{Davis {et~al}\mbox{.}(2012)Davis, Li, Mota, \&
  Winther}]{Davis_etal_2012}
Davis A.-C., Li B., Mota D.~F., Winther H.~A., 2012, ApJ, 748, 61

\bibitem[{Deffayet {et~al}\mbox{.}(2002)Deffayet, Dvali, Gabadadze, \&
  Vainshtein}]{Deffayet_etal_2002}
Deffayet C., Dvali G., Gabadadze G., Vainshtein A.~I., 2002, Phys. Rev. D, 65,
  044026

\bibitem[{{Di Matteo} {et~al}\mbox{.}(2005){Di Matteo}, {Springel}, \&
  {Hernquist}}]{DiMatteo2005}
{Di Matteo} T., {Springel} V., {Hernquist} L., 2005, \nat, 433, 604

\bibitem[{Dvali {et~al}\mbox{.}(2000)Dvali, Gabadadze, \&
  Porrati}]{Dvali_Gabadadze_Porrati_2000}
Dvali G., Gabadadze G., Porrati M., 2000, Physics Letters B, 485, 208

\bibitem[{{Farrar} \& {Peebles}(2004)}]{Farrar2004}
{Farrar} G.~R., {Peebles} P.~J.~E., 2004, \apj, 604, 1

\bibitem[{Feng {et~al}\mbox{.}(2005)Feng, Wang, \&
  Zhang}]{Feng_Wang_Zhang_2005}
Feng B., Wang X.-L., Zhang X.-M., 2005, Physics Letters B, 607, 35

\bibitem[{Gasperini {et~al}\mbox{.}(2002)Gasperini, Piazza, \&
  Veneziano}]{Gasperini_Piazza_Veneziano_2002}
Gasperini M., Piazza F., Veneziano G., 2002, Phys. Rev. D, 65, 023508

\bibitem[{Green \& Wald(2011)}]{Green_Wald_2011}
Green S.~R., Wald R.~M., 2011, Phys. Rev. D, 83, 084020

\bibitem[{{Guillet} \& {Teyssier}(2011)}]{Guillet2011}
{Guillet} T., {Teyssier} R., 2011, Journal of Computational Physics, 230, 4756

\bibitem[{Hill {et~al}\mbox{.}(2008)Hill {et~al.}}]{HETDEX}
Hill G.~J., {et~al.}, 2008, ASP Conf. Ser., 399, 115

\bibitem[{Hinterbichler \& Khoury(2010)}]{Hinterbichler_Khoury_2010}
Hinterbichler K., Khoury J., 2010, Phys. Rev. Lett., 104, 231301

\bibitem[{{Hu} \& {Sawicki}(2007)}]{Hu2007}
{Hu} W., {Sawicki} I., 2007, \prd, 76, 064004

\bibitem[{Ivezic {et~al}\mbox{.}(2008)Ivezic {et~al.}}]{LSST}
Ivezic Z., {et~al.}, 2008, arXiv:0805.2366

\bibitem[{{Jennings} {et~al}\mbox{.}(2012){Jennings}, {Baugh}, {Li}, {Zhao}, \&
  {Koyama}}]{Jennings_etal_2012}
{Jennings} E., {Baugh} C.~M., {Li} B., {Zhao} G.-B., {Koyama} K., 2012, \mnras,
  425, 2128

\bibitem[{Kaiser {et~al}\mbox{.}(2002)Kaiser {et~al.}}]{PanStarrs}
Kaiser N., {et~al.}, 2002, Proc. SPIE Int. Soc. Opt. Eng., 4836, 154

\bibitem[{{Katz} {et~al}\mbox{.}(1996){Katz}, {Weinberg}, \&
  {Hernquist}}]{Katz1996}
{Katz} N., {Weinberg} D.~H., {Hernquist} L., 1996, \apjs, 105, 19

\bibitem[{Khoury \& Weltman(2004)}]{Khoury_Weltman_2004}
Khoury J., Weltman A., 2004, Phys. Rev. D, 69, 044026

\bibitem[{Khoury \& Wyman(2009)}]{Khoury_Wyman_2009}
Khoury J., Wyman M., 2009, Phys. Rev. D, 80, 064023

\bibitem[{Kolb {et~al}\mbox{.}(2006)Kolb, Matarrese, \&
  Riotto}]{Kolb_Matarrese_Riotto_2006}
Kolb E.~W., Matarrese S., Riotto A., 2006, New J. Phys., 8, 322

\bibitem[{{Komatsu} {et~al}\mbox{.}(2011){Komatsu} {et~al.}}]{Komatsu2011}
{Komatsu} E., {et~al.}, 2011, \apjs, 192, 18

\bibitem[{Lam {et~al}\mbox{.}(2012)Lam, Nishimichi, Schmidt, \&
  Takada}]{Lam_etal_2012}
Lam T.~Y., Nishimichi T., Schmidt F., Takada M., 2012, Phys. Rev. Lett., 109,
  051301

\bibitem[{{Laureijs} {et~al}\mbox{.}(2011){Laureijs}, {Amiaux}, {Arduini},
  {Augu{\`e}res}, {Brinchmann}, {Cole}, {Cropper}, {Dabin}, {Duvet}, {Ealet},
  \& et~al.}]{EUCLID-r}
{Laureijs} R. {et~al.}, 2011, arXiv:1110.3193

\bibitem[{{Lee} {et~al}\mbox{.}(2013){Lee}, {Zhao}, {Li}, \&
  {Koyama}}]{Lee_etal_2012}
{Lee} J., {Zhao} G.-B., {Li} B., {Koyama} K., 2013, \apj, 763, 28

\bibitem[{{Li} {et~al}\mbox{.}(2012{\natexlab{a}}){Li}, {Zhao}, \&
  {Koyama}}]{Li_Zhao_Koyama_2011}
{Li} B., {Zhao} G.-B., {Koyama} K., 2012{\natexlab{a}}, \mnras, 421, 3481

\bibitem[{{Li} {et~al}\mbox{.}(2012{\natexlab{b}}){Li}, {Zhao}, {Teyssier}, \&
  {Koyama}}]{Li2012}
{Li} B., {Zhao} G.-B., {Teyssier} R., {Koyama} K., 2012{\natexlab{b}}, \jcap,
  1, 51

\bibitem[{{Llinares} \& {Mota}(2013{\natexlab{a}})}]{Llinares_Mota_2013}
{Llinares} C., {Mota} D.~F., 2013{\natexlab{a}}, Phys. Rev. Lett., 110, 161101

\bibitem[{{Llinares} \& {Mota}(2013{\natexlab{b}})}]{Llinares_Mota_2012}
{Llinares} C., {Mota} D.~F., 2013{\natexlab{b}}, Phys. Rev. Lett., 110, 151104

\bibitem[{{Lombriser} {et~al}\mbox{.}(2012{\natexlab{a}}){Lombriser}, {Koyama},
  {Zhao}, \& {Li}}]{Lombriser_etal_2012}
{Lombriser} L., {Koyama} K., {Zhao} G.-B., {Li} B., 2012{\natexlab{a}}, \prd,
  85, 124054

\bibitem[{{Lombriser} {et~al}\mbox{.}(2012{\natexlab{b}}){Lombriser},
  {Schmidt}, {Baldauf}, {Mandelbaum}, {Seljak}, \& {Smith}}]{Lombriser2012}
{Lombriser} L., {Schmidt} F., {Baldauf} T., {Mandelbaum} R., {Seljak} U.,
  {Smith} R.~E., 2012{\natexlab{b}}, \prd, 85, 102001

\bibitem[{{McQuinn} {et~al}\mbox{.}(2009){McQuinn}, {Lidz}, {Zaldarriaga},
  {Hernquist}, {Hopkins}, {Dutta}, \& {Faucher-Gigu{\`e}re}}]{McQuinn2009}
{McQuinn} M., {Lidz} A., {Zaldarriaga} M., {Hernquist} L., {Hopkins} P.~F.,
  {Dutta} S., {Faucher-Gigu{\`e}re} C.-A., 2009, \apj, 694, 842

\bibitem[{Mustapha {et~al}\mbox{.}(1997)Mustapha, Hellaby, \&
  Ellis}]{Mustapha_Hellaby_Ellis_1997}
Mustapha N., Hellaby C., Ellis G., 1997, MNRAS, 292, 817

\bibitem[{Nicolis {et~al}\mbox{.}(2009)Nicolis, Rattazzi, \&
  Trincherini}]{Nicolis_Rattazzi_Trincherini_2009}
Nicolis A., Rattazzi R., Trincherini E., 2009, Phys. Rev. D, 79, 064036

\bibitem[{{Oyaizu}(2008)}]{Oyaizu2008}
{Oyaizu} H., 2008, \prd, 78, 123523

\bibitem[{Perlmutter {et~al}\mbox{.}(1999)Perlmutter
  {et~al.}}]{Perlmutter_etal_1999}
Perlmutter S., {et~al.}, 1999, ApJ, 517, 565

\bibitem[{{Puchwein} {et~al}\mbox{.}(2005){Puchwein}, {Bartelmann}, {Dolag}, \&
  {Meneghetti}}]{Puchwein_etal_2005}
{Puchwein} E., {Bartelmann} M., {Dolag} K., {Meneghetti} M., 2005, \aap, 442,
  405

\bibitem[{{Puchwein} {et~al}\mbox{.}(2012){Puchwein}, {Pfrommer}, {Springel},
  {Broderick}, \& {Chang}}]{Puchwein2012}
{Puchwein} E., {Pfrommer} C., {Springel} V., {Broderick} A.~E., {Chang} P.,
  2012, \mnras, 423, 149

\bibitem[{{Puchwein} {et~al}\mbox{.}(2008){Puchwein}, {Sijacki}, \&
  {Springel}}]{Puchwein2008}
{Puchwein} E., {Sijacki} D., {Springel} V., 2008, \apjl, 687, L53

\bibitem[{{Puchwein} \& {Springel}(2013)}]{Puchwein2013}
{Puchwein} E., {Springel} V., 2013, \mnras, 428, 2966

\bibitem[{{Puchwein} {et~al}\mbox{.}(2010){Puchwein}, {Springel}, {Sijacki}, \&
  {Dolag}}]{Puchwein2010}
{Puchwein} E., {Springel} V., {Sijacki} D., {Dolag} K., 2010, \mnras, 406, 936

\bibitem[{Rasanen(2011)}]{Rasanen_2011}
Rasanen S., 2011, Class. Quant. Grav., 28, 164008

\bibitem[{Ratra \& Peebles(1988)}]{Ratra_Peebles_1988}
Ratra B., Peebles P. J.~E., 1988, Phys. Rev. D, 37, 3406

\bibitem[{Riess {et~al}\mbox{.}(1998)Riess {et~al.}}]{Riess_etal_1998}
Riess A.~G., {et~al.}, 1998, AJ, 116, 1009

\bibitem[{Schmidt {et~al}\mbox{.}(1998)Schmidt {et~al.}}]{Schmidt_etal_1998}
Schmidt B.~P., {et~al.}, 1998, ApJ, 507, 46

\bibitem[{Schmidt {et~al}\mbox{.}(2009)Schmidt, Lima, Oyaizu, \&
  Hu}]{Schmidt_etal_2009}
Schmidt F., Lima M.~V., Oyaizu H., Hu W., 2009, Phys. Rev. D, 79, 083518

\bibitem[{Sefusatti \& Vernizzi(2011)}]{Sefusatti_Vernizzi_2011}
Sefusatti E., Vernizzi F., 2011, JCAP, 1103, 047

\bibitem[{{Semboloni} {et~al}\mbox{.}(2011){Semboloni}, {Hoekstra}, {Schaye},
  {van Daalen}, \& {McCarthy}}]{Semboloni2011}
{Semboloni} E., {Hoekstra} H., {Schaye} J., {van Daalen} M.~P., {McCarthy}
  I.~G., 2011, \mnras, 417, 2020

\bibitem[{{Sijacki} {et~al}\mbox{.}(2007){Sijacki}, {Springel}, {di Matteo}, \&
  {Hernquist}}]{Sijacki2007}
{Sijacki} D., {Springel} V., {di Matteo} T., {Hernquist} L., 2007, \mnras, 380,
  877

\bibitem[{Sotiriou \& Faraoni(2010)}]{Sotiriou_Faraoni_2010}
Sotiriou T.~P., Faraoni V., 2010, Rev. Mod. Phys., 82, 451

\bibitem[{{Springel}(2005)}]{Springel2005c}
{Springel} V., 2005, \mnras, 364, 1105

\bibitem[{{Springel} \& {Hernquist}(2002)}]{Springel2002}
{Springel} V., {Hernquist} L., 2002, \mnras, 333, 649

\bibitem[{{Springel} \& {Hernquist}(2003)}]{Springel2003}
{Springel} V., {Hernquist} L., 2003, \mnras, 339, 289

\bibitem[{{Stanek} {et~al}\mbox{.}(2009){Stanek}, {Rudd}, \&
  {Evrard}}]{Stanek_Rudd_Evrard_2009}
{Stanek} R., {Rudd} D., {Evrard} A.~E., 2009, \mnras, 394, L11

\bibitem[{Starobinsky(1980)}]{Starobinsky_1980}
Starobinsky A.~A., 1980, Physics Letters B, 91, 99

\bibitem[{Tomita(2001)}]{Tomita_2001}
Tomita K., 2001, MNRAS, 326, 287

\bibitem[{Vainshtein(1972)}]{Vainshtein_1972}
Vainshtein A., 1972, Physics Letters B, 39, 393

\bibitem[{{van Daalen} {et~al}\mbox{.}(2011){van Daalen}, {Schaye}, {Booth}, \&
  {Dalla Vecchia}}]{VanDaalen2011}
{van Daalen} M.~P., {Schaye} J., {Booth} C.~M., {Dalla Vecchia} C., 2011,
  \mnras, 415, 3649

\bibitem[{{Viel} {et~al}\mbox{.}(2004){Viel}, {Haehnelt}, \&
  {Springel}}]{Viel2004}
{Viel} M., {Haehnelt} M.~G., {Springel} V., 2004, \mnras, 354, 684

\bibitem[{Wetterich(1988)}]{Wetterich_1988}
Wetterich C., 1988, Nuclear Physics B, 302, 668

\bibitem[{Wetterich(1995)}]{Wetterich_1995}
Wetterich C., 1995, A\&A, 301, 321

\bibitem[{Will(2005)}]{Will_2005}
Will C.~M., 2005, Living Rev. Rel., 9, 3

\bibitem[{Wiltshire(2007)}]{Wiltshire_2007}
Wiltshire D.~L., 2007, arXiv:0712.3984

\bibitem[{{Winther} {et~al}\mbox{.}(2012){Winther}, {Mota}, \&
  {Li}}]{Winther_Mota_Li_2011}
{Winther} H.~A., {Mota} D.~F., {Li} B., 2012, \apj, 756, 166

\bibitem[{Zhao {et~al}\mbox{.}(2011)Zhao, Li, \& Koyama}]{Zhao_Li_Koyama_2011a}
Zhao G.-B., Li B., Koyama K., 2011, Phys. Rev. D, 83, 044007

\bibitem[{{Zhao} {et~al}\mbox{.}(2010){Zhao}, {Macci{\`o}}, {Li}, {Hoekstra},
  \& {Feix}}]{Zhao_etal_2010}
{Zhao} H., {Macci{\`o}} A.~V., {Li} B., {Hoekstra} H., {Feix} M., 2010, \apjl,
  712, L179

\bibitem[{Zumalacarregui {et~al}\mbox{.}(2012)Zumalacarregui, Garcia-Bellido,
  \& Ruiz-Lapuente}]{Zumalacarregui_Garcia-Bellido_Ruiz-Lapuente_2012}
Zumalacarregui M., Garcia-Bellido J., Ruiz-Lapuente P., 2012, JCAP, 1210, 009

\end{thebibliography}
\end{document}